\documentclass[fleqn,10pt]{wlscirep}

\usepackage[T1]{fontenc}
\usepackage[utf8]{inputenc}

\usepackage{graphicx}
\usepackage{mathtools}
\usepackage{booktabs} 
\usepackage{textcomp} 
\usepackage{amssymb} 
\usepackage{mhchem}

\title{Observation of a Correlation Between Internal friction and Urbach Energy in Amorphous Oxides Thin Films}

\author[1,*]{Alex Amato}
\author[2]{Silvana Terreni}
\author[1]{Massimo Granata}
\author[1]{Christophe Michel}
\author[1]{Benoit Sassolas}
\author[1]{Laurent Pinard}
\author[2,3]{Maurizio Canepa}
\author[4]{Gianpietro Cagnoli}

\affil[1]{Laboratoire des Mat\'eriaux Avanc\'es, CNRS/IN2P3, F-69622 Villeurbanne (FR).}
\affil[2]{OPTMATLAB, Dipartimento di Fisica, Universit\`a di Genova, Via Dodecaneso 33, 16146 Genova, Italy.}
\affil[3]{INFN, Sezione di Genova, Via Dodecaneso 33, 16146 Genova, Italy.}
\affil[4]{Universit\'e de Lyon, Universit\'e Claude Bernard Lyon 1, CNRS, Institut Lumi\`ere Mati\`ere, F-69622, VILLEURBANNE, France.}

\affil[*]{a.amato@lma.in2p3.fr}


\begin{abstract}
We have investigated by spectroscopic ellipsometry (SE, $190-1700$ nm) the optical properties of uniform, amorphous thin films of \ce{Ta2O5} and \ce{Nb2O5} as deposited and after annealing, and after so-called “doping” with Ti atoms which leads to mixed oxides. \ce{Ta2O5} and Ti:\ce{Ta2O5} are currently used as high-index components in Bragg reflectors for Gravitational Wave Detectors. Parallel to the optical investigation, we measured the mechanical energy dissipation of the same coatings, through the so-called “loss angle” $\phi = Q^{-1}$, which quantifies the energy loss in materials. By applying the well-known Cody-Lorentz model in the analysis of SE data we have been able to derive accurate information on the fundamental absorption edge through important parameters related to the electronic density of states, such as the optical gap ($E_g$) and the energy width of the exponential Urbach tail (the Urbach energy $E_U$). We have found that $E_U$ is neatly reduced by suitable annealing as is also perceptible from direct inspection of SE data. Ti-doping also points to a minor decrease of $E_U$. The reduction of $E_U$ parallels a lowering of the mechanical losses quantified by the loss angle $\phi$. The correlation highlights that both the electronic states responsible of Urbach tail and the internal friction are sensitive to a self-correlation of defects on a medium-range scale, which is promoted by annealing and in our case, to a lesser extent, by doping. These observations may contribute to a better understanding of the relationship between structural and mechanical properties in amorphous oxides.
\end{abstract}

\begin{document}

\flushbottom
\maketitle
%
%
\thispagestyle{empty}

\section{Introduction}
\label{intro}
High quality optical interference coatings (OIC) formed by alternate low- and high-refractive index layers find scientific and technological applications in many fields of optics where the transmission or reflection of light needs to be perfectly defined as a function of wavelength. In particular in this work we are dealing with OIC that are used in low noise devices such as laser frequency stabilization cavities\cite{KesslerT.2012}, opto-mechanical resonators\cite{RevModPhys.86.1391}, atomic clocks\cite{RevModPhys.87.637} and Gravitational Wave Detectors (GWD) to name a few. 
Amorphous Oxide (AO) multilayers deposited by ion beam sputtering (IBS) are appealing for near infrared (NIR) interferometry as they are highly isotropic, homogeneous and endowed with extremely low optical absorption, which is a factor of obvious success for the best functionality of mirrors. Notably, in gravitational-wave astronomy, AO multilayers represent the key elements of high performance Bragg reflectors which are exploited in the Fabry-Perot cavities of wide-area, giant interferometer detectors (GWD)\cite{0264-9381-32-7-074001,0264-9381-32-2-024001,PhysRevD.88.043007}.
Indeed, multilayers made of Ti:\ce{Ta2O5} (the high-index material) and silica (\ce{SiO2}, the low-index material) deposited on massive, large-area fused silica substrates have been adopted by Advanced LIGO (aLIGO) and Advanced Virgo (AdV) detectors\cite{Pinard_17}, contributing to the recent, first observations of gravitational waves\cite{PhysRevLett.116.061102}.
For these GWD applications, in addition to the highest optical quality, very low internal friction is the necessary requirement to have low thermal noise \cite{Harry_2002}. In these devices noise comes from thermally activated relaxations inside the amorphous materials forming the OIC. In this respect, high-index materials (AO of transition metals like Ti and Ta) were early recognized as the dominant source of thermal noise in GWD\cite{Crooks_2004}, which limits the performance in the most sensitive region of the detection band (30 Hz to 300 Hz) where GWs originated by coalescing binaries have the strongest amplitude.
Fostering higher sensitivity in the next generation of GWD detectors, necessary to investigate deeper portions of the universe, calls for a lowering of coating thermal noise and a better knowledge of the structure of AO coatings at the molecular level\cite{Punturo_2010}.
In particular, thermal noise is associated with thermally activated transitions between equilibrium configurations of structure. There is interest to find which static properties of the amorphous structure are able to affect either the rate or the total number of such transitions. 
In the context of optical measurements, so called Urbach tails\cite{PhysRev.92.1324}, routinely observed by optical absorption measurements in crystalline and amorphous semiconductors\cite{PhysRevLett.47.1480}, describe a sub-gap exponential broadening of the absorption edge that is related to structural and thermal disorder.
The concept of band-tails states finds interesting application in the study of amorphous solids \cite{PhysRevLett.80.1928,Drabold2009} where the structural disorder is dominant with respect to thermal one. In this view, the structural origin of the exponential tails\cite{PhysRevLett.100.206403,doi:10.1002/pssa.200982877} can give precious insight in atomic organization of the system.
In this paper we present the first experimental evidence of a correlation between the amount of mechanical losses in AO wide band-gap semiconductors and the energy extension of Urbach tails. In order to show this correlation we have combined mechanical and spectroscopic ellipsometry (SE) measurements on the same coating samples.
We have considered films grown using the same procedures adopted to produce the Bragg reflectors used in aLIGO and AdV interferometers. Tantala (\ce{Ta2O5}), titania-doped tantala (Ti:\ce{Ta2O5}), niobia (\ce{Nb2O5}) and niobia-doped titania (Nb:\ce{TiO2}) amorphous thin films have been tested in their mechanical and optical response and the effect of standard annealing procedures and mixing on the optical and mechanical properties investigated. 
A possible explanation of the correlation between mechanical and optical properties may come from atomic organizations involving several tens of atoms, linked to electronic states of Urbach tail. The spatial extension of these atomic ensembles is typical of the configurations that causes internal friction and noise in amorphous \ce{Ta2O5} at room temperature\cite{Trinastic_2016}. The correlation is observed analysing different oxide coating materials, suggesting a general rather than specific validity of such properties as also suggested by the conclusions of a recent paper on Zr-doped tantala\cite{PhysRevLett.123.045501}. 

\section{Materials and methods}
Samples are IBS AO mono-layers deposited at Laboratoire des Mat\'eriaux Avanc\'es (LMA). In particular, \ce{Ta2O5} and Ti:\ce{Ta2O5} have been deposited using a custom made coater machine developed to deposit the mirrors for GWD, the so-called Grand Coater (GC)\cite{GranCoater}. The Ti-doping concentration has been optimized in order to minimize the mechanical loss\cite{Harry_2006,Amato_2019}. The atomic ratio of Ti to Ta is equal to 0.27 as measured through Rutherford backscattering spectrometry (RBS). More information about these coatings can be found elsewhere\cite{PhysRevD.93.012007}. 

\ce{Nb2O5} and Nb:\ce{TiO2} with atomic ratio of Nb to Ti equal to 0.37, as measured through energy-dispersive X-ray spectroscopy (EDX), have been deposited by a smaller custom-designed IBS machine. Niobia based materials have been investigated as candidate to replace the high-index material in GWD coatings.

Mechanical investigation requires substrates having lower internal friction than coating materials. On the other hand, optical analysis requires substrates suitable for reflection measurement. Thus, silica disk-shaped resonators (2" or 3" of diameter and 1 mm thick) have been adopted because of their low loss angle at room temperature for mechanical investigation. One-side polished silicon wafers (2" of diameter and 1 mm thick) have been used for optical investigation. 
Using the GC it was possible to deposit the same coating on the different substrates, for optical and mechanical characterization, at the same time. Using the smaller IBS machine particular attention has been dedicated to deposit coating materials in the strictly same conditions.

Coatings for GWD undergo a post-deposition heating treatment in order to reduce the optical absorption at 1064 nm and the mechanical loss. For this reason, all the samples have been analysed before and after the annealing. In particular, \ce{Ta2O5} and Ti:\ce{Ta2O5} have been annealed at 500\textcelsius~for 10 hours, following the same procedure for GWD mirrors. The crystallization temperature for \ce{TiO2} and \ce{Nb2O5} coatings is in the $250-300$\textcelsius~range and above 400\textcelsius, respectively. Instead the Nb:\ce{TiO2} coating crystallizes above 500\textcelsius. The measurements we present on \ce{Nb2O5} and Nb:\ce{TiO2} have been obtained before and after a post-deposition annealing (10 hours) at 400\textcelsius~and 500\textcelsius, respectively. 

\subsection{Mechanical Characterization}
The mechanical dissipation of amorphous coatings, related to the relaxation process, has been measured by the resonance method\cite{nowick1972anelastic} using the Gentle Nodal Suspension\cite{doi:10.1063/1.3124800,GranataInternalFriction,PhysRevD.93.012007}. In particular, the ringdown time $\tau_k$ of several vibrational modes $k$ with frequency $\nu_k$, before and after the coating deposition has been measured in order to obtain the loss angle $\phi_k=(\pi\nu_k\tau_k)^{-1}$ that quantifies the internal friction of the sample. Considering a harmonic excitation of the system, at each cycle the dissipated energy reads
\begin{equation}
E_{diss}=2\pi\phi E_{st}\medspace,
\label{eq:lossangle}
\end{equation}
where $E_{st}$ is the stored energy in the system. In this sense, equation \ref{eq:lossangle} makes explicit the relation between the loss angle and mechanical dissipation of the system.
The coating loss angle is worked out through the subtraction of the measurements of the bare substrate from that of the coated sample as detailed in ref.\cite{PhysRevD.93.012007,CAGNOLI20182165}.

\subsection{Optical Characterization}
In the purpose of finding a possible link of atomic organization to mechanical properties, an accurate investigation of the optical absorption edge in the near UV region seems promising, since it is primarily determined by the electronic joint density of states (JDOS) in turn directly depending on structural and morphological properties.
We have exploited broad-band spectroscopic ellipsometry (SE) to obtain information on the absorption edge of AO coatings.
A rotating analyzer ellipsometer (JA Woollam VASE) allowed measurements at 910 different wavelength in the  $1.1-6.5$ eV ($190-1100$ nm) energy range. The measurement of the ellipsometric angles $\Psi$ and $\Delta$\cite{fujiwara2007spectroscopic}, has been performed at different angles 55\textdegree, 60\textdegree, 65\textdegree, close to the Brewster angle $\theta_B$ (for example, $\theta_B\sim63.7\textdegree$ for \ce{Ta2O5}).
In AO, the absorption coefficient describing the optical transitions between an initial localized state in the valence band tail and an extended state in the conduction band reads 
\begin{equation}
\alpha(E) \propto e^{(E-E_0)/E_U}\medspace,
\label{eq:apha}
\end{equation} 
where $E_0$ is the energy limit of the extended band-like states and the Urbach energy $E_U$ characterizes the energy spread of the tail decay into the gap due to lattice disorder.

\section{Results}
\subsection{Mechanical Response}
\label{sec:mec}
\begin{figure*}
	\centering
	\includegraphics[width=0.95\textwidth]{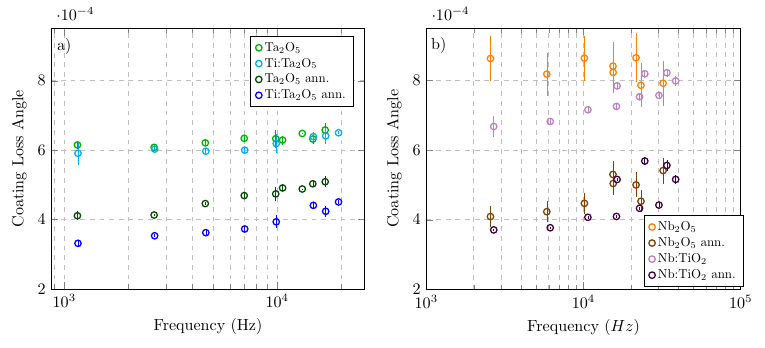}
	\caption{Coating loss angle as function of frequency for \ce{Ta2O5} (blue) and Ti:\ce{Ta2O5} (green) coating before and after the annealing on the left, for \ce{Nb2O5} (orange) and Nb:\ce{TiO2} (violet) on the right.}
	\label{fig:phi}
\end{figure*}

Figure \ref{fig:phi} shows the coating loss angle of the analysed materials, for several resonant modes in the $10^3-10^4$ Hz band.
In particular, in panel a), data taken on \ce{Ta2O5} and Ti:\ce{Ta2O5} samples show that both the annealing and the so-called “doping” treatment contribute to reduce the coating loss angle. This behaviour is in agreement with previous studies on Ti doping of annealed Ti:\ce{Ta2O5}/\ce{SiO2} and Ti:\ce{Ta2O5} coatings\cite{Harry_2006,PhysRevD.93.012007,Amato_2019}. In panel b), measurements on \ce{Nb2O5} and Nb:\ce{TiO2} are shown. The data confirm a loss angle lowering by annealing and doping.
Furthermore, the weak frequency dependence of the coating loss angle has been considered\cite{1742-6596-957-1-012006,arXiv:1909.03737} and for each sample the value of the loss angle at 10kHz is reported in table \ref{tab:correlation}.

The origin of loss can be ascribed to the peculiar structure of AO, that presents an enormous number of equilibrium configurations. The study of these equilibrium configurations is based on a pure theoretical representation known as two-level system (TLS), where two equilibrium states are described by an asymmetric double-well potential (ADWP)\cite{doi:10.1080/01418638108222343}. In order to represent the complexity of the material, different TLS having a distribution of the ADWP potential $V$ and asymmetry $\Delta$ are considered\cite{doi:10.1080/01418638108222343}. It is possible to show that at room temperature the internal friction in a material is proportional to the density of TLS that have a barrier height of about 0.5 eV\cite{Granata_2018}. In fact, following the Arrhenius law, the TLS with lower barrier height are too fast with respect to the period of the external excitation and in this way their contribution to the energy dissipation is negligible. The same happens for the TLS that have a higher barrier height for which the relaxation is much longer than the excitation period\cite{doi:10.1080/01418638108222343}.

\subsection{Optical Measurements}
\begin{figure*}[h]
	\centering
	\includegraphics[width=0.95\textwidth]{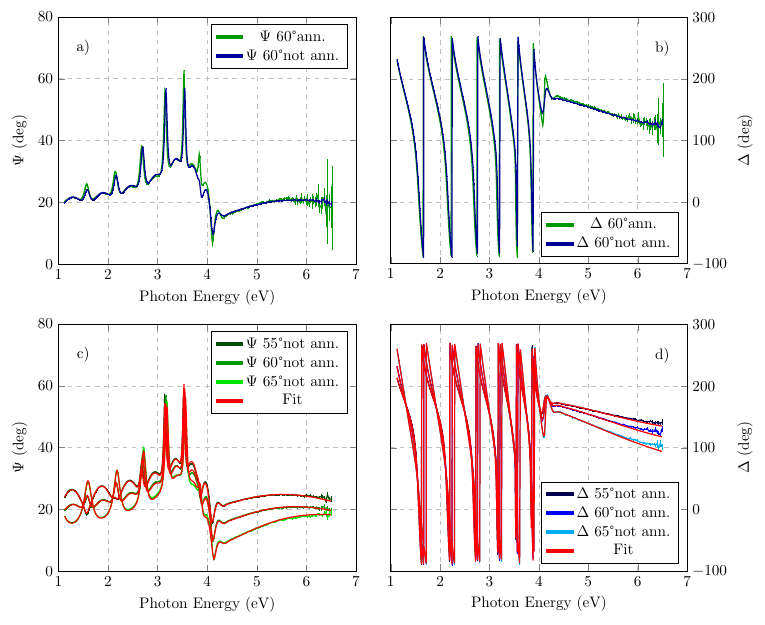}
	\caption{SE data for Ti:\ce{Ta2O5} coating. Comparison of $\Psi$ (a) and $\Delta$ (b) for measurements acquired with an angle of incident of 60\textdegree, before (blue) and after (green) the annealing. $\Psi$ (c) and $\Delta$ (d) data acquired at different angles before the annealing compared to the CL model simulation (red).}
	\label{fig:SEdata}
\end{figure*}

Considering Ti:\ce{Ta2O5} coating as a representative sample, figures \ref{fig:SEdata},a) and \ref{fig:SEdata},b) show a comparison between ($\Psi$,$\Delta$) data before and after the heating treatment.
As shown in the figure, it is clear that SE measurements before and after the annealing present evident differences in the energy region from 3.5 $eV$ to 4.5 $eV$, where the optical absorption threshold is located. In particular, after the annealing there are oscillations that are more evident near the absorption edge. As pointed out in a recent work on related systems\cite{Amato_JVST2019}, this is related to the reduction of optical transitions involving energy states in the energy gap with the annealing. We note that the same considerations can be done by comparing \ce{Ta2O5} and Ti:\ce{Ta2O5} coatings, taking into account that the energy gap of the two materials is different.

The SE data were analysed by a step by step procedure outlined below. 
\begin{itemize}
\item We have determined the optical properties of the substrate by dedicated measurements on bare samples. Creating a model that reproduces properly the optical properties of the substrate is extremely important for an accurate determination of the optical properties of coating\cite{fujiwara2007spectroscopic,MAGNOZZI201894,doi:10.1063/1.4936126}. The substrate data have been analysed by the WVASE GenOsc routine considering an ultra-thin oxide (about 2 nm resulting from fit); some details on this analysis was included in ref.\cite{Amato_2019}.

\item We have fitted the substrate+coating data in the transparency range of the coating with interpolation formula (Cauchy, Sellmeier) to obtain a first reliable estimate of NIR refractive index and coating thickness (as done in ref.\cite{prato2011gravitational}).

\item The obtained coating thickness has been used in the so-called point-by-point analysis which provided a first determination of the coating dielectric function in the whole measured energy range.

\item The dielectric function obtained by the point-by-point analysis has been used to check results obtained by application of a convenient parametric model, including Urbach absorption tails. The optical parameters and the thickness of the coating were then determined by fitting the $(\Psi, \Delta)$ data\cite{fujiwara2007spectroscopic}. 
\end{itemize}

Previous work on transition metal amorphous oxides (in ref.\cite{prato2011gravitational} and references therein), showed that the KK-consistent Cody-Lorentz (CL) formula\cite{ferlauto2002analytical} for the dielectric function $\varepsilon$ provides an adequate interpretation of optical properties in a wide energy region across the absorption edge. The CL model accounts for the Urbach tail and includes $E_U$ as a relevant parameter to be exploited in the fitting procedure. As example, in figures \ref{fig:SEdata},c) and \ref{fig:SEdata},d) the CL model simulation is compared to the ($\Psi$,$\Delta$) experimental data for Ti:\ce{Ta2O5} before the heating treatment. A Bruggeman effective medium approximation (EMA) layer has been added to account for roughness, which allows to improve the fit model especially in the strong absorption region (UV), where light mostly probes the surface features of the film. We attempted to model the possible interface by practicable models but the fit regression code usually nulled the interface thickness. The sharpness of the substrate/coating interface is also indirectly confirmed by the surface roughness which is at atomic level as obtained by atomic force microscopy (AFM) measurements of root mean square roughness on this kind of samples\cite{prato2011gravitational}.
The results obtained by the best fit with the CL model, are shown in table \ref{tab:correlation}. Figures illustrating the point-by-point analysis and the comparison with the CL model analysis can be found in the supplementary information file.

The best fit determination of the complex dielectric function allows to calculate the absorption coefficient through the well-known relation $\alpha(E) = (\varepsilon_2 E)/ (\hbar c n)$, where $\varepsilon_2$ is the imaginary part of $\varepsilon$, $n$ is the refractive index, $c$ the speed of light in [cm/s] and $\hbar$ the reduced Plank constant.

\begin{figure*}[h]
	\centering
	\includegraphics[width=0.95\textwidth]{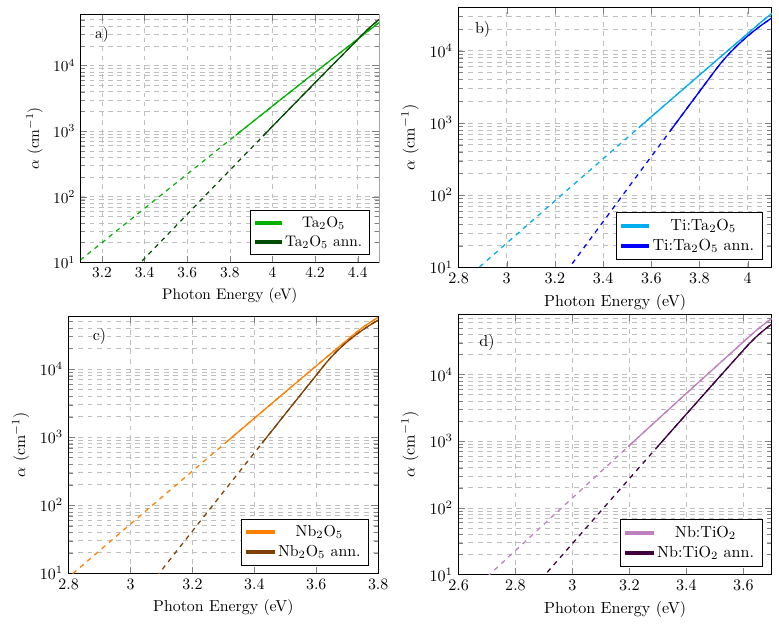}
	\caption{Absorption coefficient of \ce{Ta2O5} (top) and \ce{Nb2O5} (bottom) before and after the annealing, for undoped (left panels) and doped (right panels) material. The curves have been calculated starting from the extinction coefficient functions provided by the best fit CL model. The logarithmic scale emphasizes the exponential decay of absorption in the Urbach region. The dashes indicate regions where the sensitivity of the measurements is poor and curves should be intended as mere extrapolations of the model.}
	\label{fig:alpha}
\end{figure*}

Figure \ref{fig:alpha} shows the absorption coefficient as the function of the photon energy for the same materials analysed in previous section. In the logarithmic scale representation chosen for $\alpha$, Urbach tails have a linear appearance and the inverse of the slope is proportional to $E_U$ (equation \ref{eq:apha}).
Panels a), b), c) and d) show the effect of the annealing on \ce{Ta2O5}, Ti:\ce{Ta2O5}, \ce{Nb2O5} and Nb:\ce{TiO2} coatings, respectively. \\
In the figure, it is possible to appreciate the reduction of the extension of the Urbach tails. The reduction of the optical gap induced by doping (see Table \ref{tab:correlation}) is also evident. The values confirm the homogeneity of the coatings since, especially for \ce{Ta2O5} the shift of the energy gap due to the Ti-doping, where the \ce{TiO2} has been found to have an energy gap about $3.3-3.5$ eV\cite{doi:10.1063/1.356306,NAKAMURA2000506}, is in agreement with the atomic ratio concentration. The shift of the energy gap is less evident for Nb:\ce{TiO2} as the two oxides have similar energy gap.
For a more quantitative comparison, the values of the Urbach energy obtained by the fit using the CL model are reported in table \ref{tab:correlation}. From the table it can be observed that, as already noticed, also Ti/Ta mixing induces a reduction of $E_U$ but to an extent which is smaller than the annealing-induced variation. Anyway, the greatest reduction can be obtained by suitably combining annealing and doping.  

\subsection{Correlation between mechanical loss and Urbach energy}
\begin{figure}[h]
	\centering
	\includegraphics[width=0.41\textwidth]{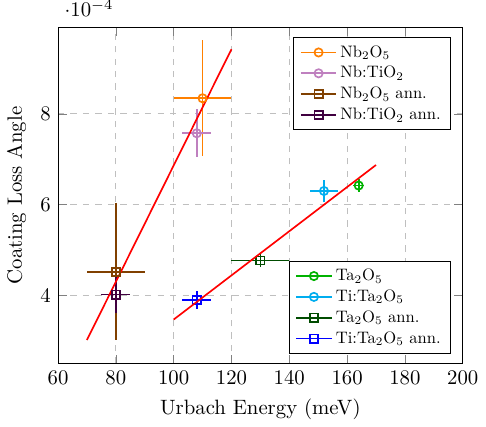}
	\caption{Coating loss angle as function of Urbach energy. The circles and the squares refer to as deposited samples and samples after the heating treatment, respectively. Four different coatings are considered, the \ce{Ta2O5} (blue) and the Ti:\ce{Ta2O5} (green), the \ce{Nb2O5} (orange) and the Nb:\ce{TiO2} (violet).}
	\label{fig:correlation}
\end{figure}

The results about the coating loss angle and the optical absorption coefficient reported in figure \ref{fig:phi} and \ref{fig:alpha} respectively, exhibit a similar trend regarding the annealing and the mixing. In this respect, it is interesting to look at figure \ref{fig:correlation} where the coating loss angle is reported as function of the Urbach energy $E_U$, for both tantala and niobia coating under different conditions. The values used in figure \ref{fig:correlation} are summarised in table \ref{tab:correlation}. The figure clearly suggests the existence of a positive correlation between mechanical losses and extension of the Urbach tail. 
\begin{table}
	\centering
	\caption{Most relevant parameters of the mechanical and optical analysis of coating materials before and after the annealing. The reported coating loss angle $\phi_c$ is the value at 10 kHz obtained by applying to each data set shown in figure \ref{fig:phi} a least-square regression fitting with a frequency-dependent loss.  $E_U$ and $E_g$ are the Urbach energy and Energy gap, respectively. The fit uncertainty on optical parameters and the Mean Square Error (MSE) are provided by the regression code (WVASE).}
	\label{tab:correlation}
	\begin{tabular}{lccccc}
		\hline
		\noalign{\smallskip}
		Coating                              & $\phi_c$ @ 10 kHz & Thickness &   $E_U$    &    $E_g$     & MSE \\
		                                                            &  $\cdot10^{-4}$   &   (nm)    &   (meV)    &     (eV)     & \\
		\noalign{\smallskip}\hline\noalign{\smallskip}
		\ce{Ta2O5} 													&   $6.42\pm0.15$   & $579\pm2$ & $164\pm8$  & $4.1\pm0.2$  &4.0\\
		Ti:\ce{Ta2O5}                                               &    $6.3\pm0.2$    & $500\pm2$ & $152\pm5$  & $3.6\pm0.1 $ &4.1\\
		\ce{Ta2O5} ann.                                             &   $4.77\pm0.15$   & $592\pm2$ & $130\pm10$ & $4.0\pm0.1 $ &11.0 \\
		Ti:\ce{Ta2O5} ann.                                          &    $3.9\pm0.2$    & $509\pm2$ & $108\pm5$  & $3.6\pm0.1 $ &5.2\\
		\noalign{\smallskip}\hline\noalign{\smallskip}
		\ce{Nb2O5}													&    $8.3\pm1.3$    & $470\pm2$ & $110\pm10$ & $3.4\pm0.1 $ &5.9\\
		Nb:\ce{TiO2}                                                &    $7.6\pm0.5$    & $482\pm2$ & $108\pm5$  & $3.3\pm0.1 $ &7.7\\
		\ce{Nb2O5} ann.                                             &    $4.5\pm1.5$    & $483\pm2$ & $80\pm10$  & $3.4\pm0.1 $ &7.1\\
		Nb:\ce{TiO2} ann.                                           &    $4.0\pm0.4$    & $490\pm2$ &  $80\pm5$  & $3.3\pm0.1 $ &7.6\\
		\noalign{\smallskip}\hline                                  & 
	\end{tabular}
\end{table}

To our knowledge, this is the first time that such correlation is reported.

\section{Discussion}
The observed correlation between coating loss angle and Urbach energy calls for a common physical background, to be most naturally searched in the structural ground.
In this respect interesting results were obtained on tantala and Ti-doped tantala coatings, i.e. the same system investigated here (though deposited in a different laboratory under different conditions), by combining loss angle and TEM measurements, corroborated by MD simulations\cite{BASSIRI20131070}.
It has been shown through Radial density function (RDF) measurements on Ti:\ce{Ta2O5} coating that a decrease of the measured loss angle is accompanied by a small improvement at the short-range scale inferred by a width reduction of the main RDF peak, related to metal-oxygen distances.
More recent studies on \ce{Ta2O5}\cite{Shyam2016} highlighted that the amorphous material is made of Primary Structural Units (PSU) very similar to those of the crystalline state. The authors claimed that annealing favours the organization of PSUs in short 1D-chains or 2D ribbons, whereas the influence of doping (Zr in that case) is to modify and perhaps suppress formation of ribbons. Further they suggest that the emergence or suppression of medium range 2D order may have important influence on mechanical properties of \ce{Ta2O5} coating. A very recent paper\cite{PhysRevLett.123.045501} on the Zr:\ce{Ta2O5} system confirms that annealing produces systematic changes at the intermediate range scale; atomic modelling shows that such changes are to be related to well definite changes of the connections between PSU. These changes in the amorphous structure correlate with a reduction of mechanical losses.

In another recent work, Raman spectroscopy measurements on \ce{SiO2} coating showed that extended structures such as rings made of three tetrahedrons of \ce{SiO2} (the PSU of \ce{SiO2}) are correlated to the coating loss angle\cite{PhysRevMaterials.2.053607} and have an activation energy of about 0.43 eV\cite{PhysRevB.48.15672}. Furthermore, molecular dynamics of AO such as \ce{SiO2} showed that TLS with a barrier of 0.5 eV primarily involves quasi-1D chain of Si-O-Si and rings of Si-O-Si bonds\cite{PhysRevB.97.014201}. The ensemble of measurements strongly suggests that at room temperature the main contribution to the loss angle is to be ascribed not only to short-range order but also to non-local structural organization.

The evidence for a relation between Urbach tails and structure is dating back to early works on the subject: Cody et al.\cite{PhysRevLett.47.1480}, recognized that the width of the exponential tail is controlled by two additive terms which represent the contribution of structural and thermal disorder in the network, respectively. More recent works\cite{PhysRevLett.100.206403,PhysRevB.83.045201} have pointed out a relation between the presence of structural, not crystalline atomic organization on a medium range scale and the extension of Urbach tails. 
Interesting results have been obtained by atomistic modelling of amorphous silicon\cite{PhysRevLett.100.206403}: molecular dynamics calculations show that after the relaxation of the structure an exponential valence tail appears in the electronic DOS. An inverse participation ratio (IPR) analysis shows that the extreme tail eigenstates on amorphous semiconductors are primary localized on so-called topological filaments (TF). 
In fact, while well localized defects induce mid-gap states, more complex and organized structures induces Urbach states near the valence or conduction band edge. In a-Si such structures are connected subnetworks of short bonds or long bonds \cite{Drabold2009}. Interesting, if the defects correlation is artificially destroyed the Urbach tail is severely affected. Structural relaxation favours defects correlation and reduces the Urbach energy\cite{PhysRevLett.100.206403}.

Post-deposition annealing generally modifies the Urbach tails. For example, Xue et al.\cite{XUE200821} recorded the decreasing of Urbach energy in ZnO thin film with increasing annealing temperature from 600\textcelsius~to 750\textcelsius, whereas an inverse trend is observed exceeding 750\textcelsius. This observation can help to understand Urbach tailing mechanisms, where the annealing may be used to allow a structural self-organization with a consequent lattice strain relaxation\cite{Boubaker2011}. 

Our measurements (at room temperature) lead us to advance the hypothesis that the structural disorder contribution to the Urbach energy decreases as result of annealing-induced organization at the medium-long range.  Further, structures which ordering spans beyond the short range are likely involved in thermal dissipation active at room temperature, but this could be no longer true at low temperatures. The correlation we observed at room temperature between Urbach energy and relaxation mechanisms is noteworthy; how things change with temperature is indeed of great scientific interest and will be the subject of future experimental investigation.

Therefore, Urbach tails can be viewed as a simple, meaningful way to visualize the occurrence of atomic organization in the amorphous structure, in a multi-range perspective. In this respect, the correlation between the Urbach tail extension and the mechanical losses could be explained through the spatial character of the Urbach tails states, related to atomic configurations that are responsible of the dissipation of mechanical energy as measured at 300 K in the acoustic band. Annealing relaxes the network and consequently this relaxation increases the spatial correlation between defects: the structure evolves from large stresses concentrated in small regions towards a situation where weakly strained regions are clustered around the site where the large stress was once. Urbach-type electronic states associated to medium/long range atomic configurations approach in energy the mobility edge so that Urbach tails get narrower and Urbach energy decreases. In order to explain the impact of stress relaxation on energy loss one has to consider that the measurement were done only at room temperature, hence, only the reduction of TLS density having a barrier height of about 0.5 eV is probed. 
Energy barriers of interest at room temperature correspond to equilibrium configurations formed by several PSUs, as suggested in a recent work\cite{Trinastic_2016,Shyam2016}. 
The recent observations of ref.\cite{PhysRevLett.123.045501} have been interpreted by the authors as due to a decrease of PSUs sharing edges in favour of corner sharing which form TLS with lower barrier heights.

The effect of Ti/Ta mixing on the reduction of the Urbach energy is less intuitive and to some extent even counter-intuitive. 
The investigation done in ref.\cite{BASSIRI20131070} points out that Ti-doping changes the \ce{Ta2O5} structure, possibly leading to an increase of short-range homogeneity. The authors showed that this is related to a low coating loss angle. In ref.\cite{BASSIRI20131070} reverse Monte Carlo (RMC) simulation has been used to match the observed RDF and a significant fraction of \ce{TaTiO2} ring fragments is formed in the doped coating. This changes the structure by modifying the angles between oxygen-metal-oxygen and metal-oxygen-metal as consequence. 
The considerations made above suggest a possible explanation of the broadening reduction with the reduction of the structural disorder contribution to $E_U$ as a consequence of both annealing (medium range atomic ordering) and mixing (short range homogeneity, as suggested in ref.\cite{BASSIRI20131070}). It is worth noting that also mixing-induced variations of the phonon spectrum can lead to a reduction of the thermal contribution to $E_U$.
In ref.\cite{Studenyak2014} such contribution was estimated through an Einstein model whose characteristic temperature $\theta_E$,  corresponding to an average phonon temperature, approximates a Debye temperature $\theta_D\simeq4\theta_E/3$. 
Therefore, a slight increase of the Debye temperature in the mixed materials could be related to the observed slight reduction of $E_u$.

In any case, since the Urbach tails of doped oxide are less extended than the undoped one, we can infer that the local structure of the doped material is more homogeneous and that on a larger scale  the material become organized into cluster of atoms resulting  in a lower loss angle.

\section{Conclusions}
The search of the structural origin of the mechanical dissipation behaviour in amorphous materials has produced several sound results in recent years. \\
In a early work\cite{BASSIRI20131070} a correlation between the mechanical coating loss angle and the short-range structural organization has been claimed. More recent studies\cite{Trinastic_2016,PhysRevB.97.014201} showed that the TLS, giving rise to the dissipation mechanism at room temperature, involve more complex structures, regarding a medium- and even long-range structural organization. This seems fully confirmed from the recent observations of ref.\cite{PhysRevLett.123.045501}\\
The Urbach energy is a parameter which probes the degree of atomic organization on a multi-range scale by optical absorption investigation. In this work, both internal friction and Urbach energy have been studied for \ce{Ta2O5} and \ce{Nb2O5} coating materials, under different condition regarding the mixing with \ce{TiO2} and a post-deposition heating treatment at 500\textcelsius~for 10 hours.\\
The annealing relaxes the structure of coating materials, leading to a more organized structure, still maintaining the amorphous state. In the same way, a certain amount of Ti doping either modifies the structure forcing the system in a more homogeneous atomic disposition or it changes the phonon spectrum reducing effectively the impact of temperature on the Urbach energy. The annealing, primarily, and to a less extent the Ti doping eventually reduce the mechanical loss angle and the Urbach energy, leading to a correlation between these quantities. \\
The observed correlation between the energy extension of Urbach tails and the level of mechanical losses opens new perspectives.  
The characterization of Urbach tails complements the analysis with purely structural methods. Being the Urbach energy a single value parameter hardly it can describe the complexity of the amorphous material, whereas it is correlated to the atomic organization of the material at the right range for the energy loss mechanisms.
The correlation we found at room temperature points to the importance of the structural organization at the medium range scale for reducing internal friction. Measurements as a function of temperature would desirable to better disentangle the two contributions to $E_U$ and monitor the relation with low frequency mechanical losses
Finally, since the correlation has been proved with three different oxides (\ce{Ta2O5} and \ce{Nb2O5} mixed with \ce{TiO2}), further investigation  could be interesting to check the validity of the correlation for amorphous systems in general.

\bibliography{bibliography}

\section*{Acknowledgements}
The authors wish to acknowledge Gianluca Gemme for his strong support to this activity and Michele Magnozzi for experimental assistance. This work has been part of the the subsystem Coating Research Development of the Advanced Virgo + project and the authors are member of the Virgo Coating R\&D collaboration.

\section*{Author contributions statement}
A.A. conceived the experiment, collaborated to the mechanical characterization of the samples, made the optical characterization of the samples, collaborated to the interpretation of spectroscopic ellipsometry data and internal friction measurements, collaborated to writing and revising the manuscript.

\noindent S.T. collaborated to the interpretation of spectroscopic ellipsometry data, collaborated to writing and revising the manuscript.

\noindent M.G. collaborated to the mechanical characterization of the samples.

\noindent C.M., B.S. and L.P. produced the coating samples.

\noindent M.C. conceived the experiment, collaborated to the interpretation of spectroscopic ellipsometry data, collaborated to writing and revising the manuscript.

\noindent G.C. supervised the mechanical measurements, collaborated to the interpretation of internal friction measurements, collaborated to writing and revising the manuscript.

\section*{Additional information}
\textbf{Competing Interests Policy:} The authors declare no competing interests.

\end{document}